\documentclass[12pt]{iopart}
\usepackage[english]{babel}
\usepackage{graphicx}
\usepackage{caption}
\usepackage{subfig}
\usepackage{epsfig}

\newcommand{\beq}{\begin{equation}}
\newcommand{\eeq}{\end{equation}}
\newcommand{\beqa}{\begin{eqnarray}}
\newcommand{\eeqa}{\end{eqnarray}}

\begin{document}

\title{Pair condensation of polarized fermions in the BCS-BEC crossover}

\author{G. Bighin$^{1,2}$, G. Mazzarella$^{1,3}$, 
L. Dell'Anna$^{1,3}$, and L. Salasnich$^{1,3,4}$}
\address{$^{1}$Dipartimento di Fisica e Astronomia ``Galileo Galilei'', 
Universit\`a di Padova, Via Marzolo 8, 35131 Padova, Italy \\
$^{2}$Istituto Nazionale di Fisica Nucleare (INFN), Sezione di Padova, 
Via Marzolo 8, 35131 Padova, Italy \\
$^{3}$Consorzio Nazionale Interuniversitario per le Scienze Fisiche 
della Materia (CNISM), Unit\`a di Padova, Via Marzolo 8, 35131 Padova, Italy\\
$^{4}$Istituto Nazionale di Ottica (INO) del Consiglio Nazionale 
delle Ricerche (CNR), Sezione di Sesto Fiorentino, 
Via Nello Carrara, 1 - 50019 Sesto Fiorentino, Italy}


\begin{abstract}
We investigate a two-component Fermi gas with unequal spin populations 
along the BCS-BEC crossover. By using the extended BCS equations 
and the concept of off-diagonal-long-range-order we derive a formula for the 
condensate number of Cooper pairs as a function of energy gap, 
average chemical potential, imbalance chemical potential and temperature. 
Then we study the zero-temperature condensate fraction 
of Cooper pairs by varying interaction strength and polarization, 
finding a depletion of the condensate fraction by increasing 
the population imbalance. We also consider explicitly 
the presence of an external 
harmonic confinement and we study, within the local-density approximation,  
the phase separation between superfluid and normal phase regions 
of the polarized fermionic cloud. 
In particular, we calculate both condensate density profiles and 
total density profiles from the inner superfluid core to the normal region 
passing for the interface, where a finite jump in the density is a clear 
manifestation of this phase-separated regime. Finally, 
we compare our theoretical results with the available experimental data 
on the condensate fraction of polarized $^6$Li atoms 
[Science {\bf 311}, 492 (2006)]. These experimental data 
are in reasonable agreement with our predictions in a suitable range 
of polarizations, but only in the BCS side of the crossover up to unitarity.
\end{abstract}

\pacs{03.75.Ss, 05.30.Fk, 67.85.Lm}

\maketitle

\section{Introduction}

The experimental realization of the predicted crossover from the 
Bardeen-Cooper-Schrieffer (BCS) state of weakly bound Fermi pairs to the 
Bose-Einstein condensate (BEC) of molecular dimers \cite{eagles,leggett,
nozieres, greiner,regal,kinast,zwierlein,chin,ueda} 
is one of the most important achievements of atomic physics 
over the past years. In two experiments 
\cite{zwierlein,ueda} the condensate fraction of Cooper pairs \cite{yang}, 
which is directly related to the off-diagonal-long-range-order of the 
two-body density matrix of fermions \cite{penrose,campbell}, has been 
investigated with two-hyperfine-component Fermi vapors of $^{6}$Li atoms in 
the BCS-BEC crossover. The experimental results exhibit a quite good agreement 
with mean-field theoretical predictions \cite{salasnich1,ortiz} and 
Monte-Carlo simulations at zero temperature \cite{monte-carlo}. 
Moreover, the condensate fraction is a relevant 
quantity also in understanding the BCS-BEC crossover 
in the presence of spin-orbit couplings \cite{so1,so2}, 
in the case of a narrow resonance \cite{sala-narrow}, and also 
for the two-dimensional Fermi 
superfluid \cite{sala-2d,sala-2dlattice,sala-beyond}. 

A very interesting extension of these studies is the analysis of two-component 
trapped Fermi gases with polarization, i.e. with a population 
imbalance between the two hyperfine components. 
Many efforts have been devoted to this topic both theoretically 
\cite{bedaque,carlson,pao0,son,miz,sheehy1,mannarelli,pieri,liu,hu,chien,
gu,martikainen,iskin2006,desilva,haque,yi,kinnuen,parish1,sheehy2} and 
experimentally \cite{zwierlein0,partridge,zwierlein1}. The polarized Fermi 
gas is characterized by a far richer phase diagram than the equal 
spin-population case. Such a physical system, in fact, exhibits a quantum 
phase transition between the superfluid (SF) and normal (N) states, and it 
has been predicted to possess exotic SF phases such as the inhomogeneous 
Fulde-Ferrell-Larkin-Ovchinnikov state \cite{fulde,larkin}, and a 
phase-separated (PS) regime \cite{bedaque,carlson,hu,gu,sheehy1,iskin2006,
haque,parish1,sheehy2,zwierlein0,partridge} where a normal phase coexists 
with a superfluid state. 
Zwierlein {\it et al} \cite{zwierlein0} have measured the condensate 
fraction of Cooper pairs 
as a function of the polarization both in the BCS region and on 
the BEC side of the crossover. However, on the theoretical side, 
it is still missing a systematic analysis of the condensate fraction 
in the imbalanced case. 

In this paper we provide a theoretical investigation of pair condensation 
of polarized fermions in the BCS-BEC crossover. 
By using a path-integral formalism we 
derive for the polarized superfluid Fermi gas an analytical formula 
for the condensate fraction of Cooper pairs, which is then studied with 
generalized BCS equations, at zero temperature and in the uniform case, 
as a function of the dimensionless interaction parameter $y=1/(k_Fa_s)$, 
with $k_F$ the Fermi wave vector and $a_s$ the s-wave fermion-fermion 
scattering length, and for different polarizations. 
From this analysis we find that an increase of the population 
imbalance gradually produces a depletion of the condensate fraction, 
which is shown to decrease linearly with the polarization on the 
BEC side of the resonance. 
Finally, we consider the inclusion of a realistic axially-symmetric 
harmonic potential. Within the local density approximation 
we analyze the behavior of both total density profile and condensate 
density profile for different scattering lengths 
finding a finite jump in the profiles, which is a clear 
signature of the phase boundary between the superfluid phase 
(in the inner core) and normal phase (in the outer region) where pair 
condensation is absent. These results are compared with the experimental 
ones \cite{zwierlein0}. 

\section{Extended BCS equations for the polarized Fermi gas}
\label{sec:ebcs}

We consider a two-spin-component ($\sigma=\uparrow,\downarrow$) Fermi gas 
with unequal spin populations having the same mass $m$. The fermions are 
supposed to interact via a contact potential. In the absence of external 
confinement, our system can be described by the following (Euclidean) 
Lagrangian density \cite{tempereft1,tempereft2}:
\beqa
\mathcal{L} &=& \sum_{\sigma = \uparrow, \downarrow} \bar{\psi}_\sigma 
\left( \mathbf{r}, \tau \right) \left( \hbar \frac{\partial}{\partial 
\tau}- \hbar^2 \frac{\nabla^2}{2m} - \mu_\sigma \right) \psi_{\sigma} 
\left( \mathbf{r}, \tau \right)\nonumber\\
&+&g \,\bar{\psi}_\uparrow \left( \mathbf{r}, \tau \right)  
\bar{\psi}_\downarrow \left( \mathbf{r}, \tau \right)  \psi_\downarrow 
\left( \mathbf{r}, \tau \right)  \psi_\uparrow \left( \mathbf{r}, \tau \right)
\; . 
\label{eq:lagr0}
\eeqa
Here $\psi_{\sigma}$, $\bar{\psi}_{\sigma}$ are the Grassmann field variables; 
$g<0$ is the strength coupling of the fermion-fermion attractive 
contact-potential interaction; $\mu_\sigma$ is the chemical potential 
of the component $\sigma$. 
In the rest of the paper for simplicity we set $\hbar=k_B=1$, 
with $\hbar$ the reduced Planck constant and $k_B$ the Boltzmann constant.  
We define the average chemical 
potential $\mu$ and the imbalance chemical potential 
$\zeta$ as the half-sum and 
the half-difference between the the two chemical potentials, respectively, i.e.
\beq
\label{hshd}
\mu = \frac{\mu_\uparrow + \mu_\downarrow}{2} \hspace{22pt} \zeta = 
\frac{\mu_\uparrow - \mu_\downarrow}{2} \; . 
\eeq
In the following, without loss of generality, we assume that the $\uparrow$ 
species is the majority component.

We are interested in studying the thermodynamics of the system 
at temperature $T$ in a volume $V$. All the thermodynamical properties 
can be inferred from the partition function $Z$,  
which in the path integral formalism reads
\beq
Z = \int \mathcal{D} \psi \mathcal{D} \bar{\psi} \ e^{-S \left[ \psi, 
\bar{\psi} \right]} \; , 
\label{eq:partfunc}
\eeq
where 
\beq
S \left[ \psi, \bar{\psi} \right] = \int_0^\beta \mathrm{d} \tau \int_V 
\mathrm{d}^3\mathbf{r} \ \mathcal{L} \; 
\eeq
is the Euclidean action with $\beta=1/T$ \cite{tempereft1,tempereft2}. 

From the partition function, one can calculate the number of fermions 
with spin $\sigma$ given by 
\beq
\label{numbersigma}
N_{\sigma} =\frac{1}{\beta} \frac{\partial}{\partial \mu_\sigma} \ln Z \; . 
\eeq
The thermodynamics of the system can be studied by decoupling the Grassmann 
field-quartic interaction term in the second row of Eq. (\ref{eq:lagr0}) by 
means of a Hubbard-Stratonovich transformation. The resulting complex auxiliary 
field $\Delta (\mathbf{r}, \tau)$ is then expanded 
around its saddle point value $\Delta (\mathbf{r}, \tau) = \Delta_0 + 
\delta (\mathbf{r}, \tau)$. At this stage we treat our system by using 
the mean-field (MF) approximation which consists in neglecting the 
fluctuations $\delta (\mathbf{r}, \tau)$. In this approximation, $\Delta_0$ 
is the real MF order parameter \cite{tempereft1,tempereft2}. 

In the Nambu-Gorkov formalism \cite{nambu}, 
our theory is described by the following 
$2 \times 2$ momentum-space Green function:
\beq
{{\widetilde G}^{-1}} = 
\left( 
\begin{array}{cc}
i \omega_n + 
\frac{k^2}{2m} - \mu - \zeta & \Delta_0 \\
{\Delta}_0 & i \omega_n - \frac{k^2}{2m} + \mu - \zeta
\end{array}
\right) \label{eq:green1}
\eeq
from which it is immediate to get the energy spectrum of the Bogoliubov 
excitations:
\beq
E^{\pm}_{\bf k} = 
\sqrt{\xi_{\bf k}^2 + \Delta_0^2} \pm \zeta \;,
\label{bogoliubov} 
\eeq
where $\xi_{\bf k} = \epsilon_{\bf k} - \mu = \frac{k^2}{2m} - \mu$ 
and $\Delta_0$ is the familiar energy gap of fermionic elementary excitations. 
The sign plus (minus) in 
Eq. (\ref{bogoliubov}) holds for $\uparrow$ ($\downarrow$) component. 
The fermionic fields $\bar{\psi} \left( \mathbf{r}, \tau \right)$ and 
$\psi \left( \mathbf{r}, \tau \right)$ can be integrated out, and the 
summation over Matsubara frequencies yields the following effective action:
\beq
S_{e} = - \beta V \frac{\Delta_0^2}{g} 
+\beta \sum_{\bf k} \xi_{\bf k} \nonumber 
- \sum_{\bf k} \ln\left(2 \cosh \left( \beta \zeta \right) + 2 
\cosh \left( \beta E_{\bf k}\right)\right) 
\label{seff}
\eeq
with $E_{\bf k} = \sqrt{\xi_{\bf k}^2 + \Delta_0^2}$. 
From the effective action (\ref{seff}) we achieve the extended BCS (EBCS) 
equations at finite temperature  \cite{tempereft1,tempereft2}:
\beq
\frac{1}{g} = \frac{1}{V} \sum_{\bf k} \frac{1}{E_{\bf k}} 
\frac{\sinh \left( \beta E_{\bf k} \right)}
{2 \cosh\left( \beta \zeta\right) + 
2 \cosh\left( \beta E_{\bf k}\right)} \; , 
\label{gapeqT}
\eeq
\beq
N=\sum_{\bf k} \left( 1- \frac{\xi_{\bf k}}{E_{\bf k}} 
\frac{\sinh \left( \beta E_{\bf k} \right)}{\cosh \left( \beta \zeta\right) 
+ \cosh \left( \beta E_{\bf k}\right)} \right) \; , 
\label{numbereqT}
\eeq
\beq
N_{\uparrow} - N_{\downarrow} 
= 2 \sum_{\bf k} \frac{\sinh \left( \beta \zeta \right)}
{\cosh \left( \beta \zeta\right) + \cosh \left( \beta E_{\bf k}\right)} 
\; . 
\label{imbeqT}
\eeq
The first of the three above equations is the gap equation, Eq. 
(\ref{numbereqT}) is the number equation, and, finally, Eq. (\ref{imbeqT}) 
is the equation for the population imbalance.

The gap equation (\ref{gapeqT}) needs to be regularized, and this can be 
done according to the following prescription (see, for instance, 
\cite{leggett}):   
\beq
\frac{m}{4 \pi a_s} = - \frac{1}{g} + \frac{1}{V} \sum_{{\bf k}} 
\frac{1}{2 \epsilon_{\bf k}} \;, 
\eeq
where $a_s$ is the s-wave scattering length.
In such a way one attains the regularized gap equation 
\beq
\frac{m}{4 \pi a_s} = \frac{1}{2 V} \sum_{\bf k} \left( \frac{1}
{\epsilon_{\bf k}} - \frac{1}{E_{\bf k}} \frac{\sinh 
\left( \beta E_{\bf k} \right)}{\cosh \left( \beta \zeta\right) + 
\cosh \left( \beta E_{\bf k}\right)} \right) \; . 
\label{gapeqTreg}
\eeq

Finally, we calculate the grand potential $$\displaystyle{\Omega=-
\frac{1}{\beta}\ln Z}.$$ By using in this equation, Eqs. (\ref{eq:partfunc}) 
and (\ref{seff}), the mean-field version of $\Omega$ is given by
\beq
\Omega = -V \frac{\Delta_0^2}{g}+\sum_{{\bf k}} 
\left( \xi_{\bf k} - E_{\bf k} \right) 
- \frac{1}{\beta}\,\sum_{{\bf k}}\big[\ln(1+e^{-\beta(E_{\bf k}-\zeta)})
+\ln(1+e^{-\beta(E_{\bf k}+\zeta)})\big] \; . 
\label{omegaT}
\eeq
In the limit of zero temperature ($T=0$, i.e. $\beta\to \infty$), 
the MF grand potential 
(\ref{omegaT}) reads
\beq
\Omega_{T=0} = 
\sum_{{\bf k}} \left( \xi_{\bf k} - E_{\bf k} 
\right) - V \frac{\Delta_0^2}{g} + 
\sum_{k_- < \left| {\bf k} \right| < k_+} 
\left( E_{\bf k} -\zeta \right) \; , 
\label{omeganoT}
\eeq
where $k_-= \sqrt{2m} \sqrt{ \max \left( \mu - \sqrt{\zeta^2 - \Delta_0^2},0
\right) } $ and $k_+ = \sqrt{2m} \sqrt{\max(\mu + \sqrt{\zeta^2 - 
\Delta_0^2},0)}$.

The first two terms of Eq. (\ref{omeganoT}) are the same 
as in the unpolarized two-component Fermi gas, while the last one describes 
the polarized case and contributes to the physics of the system provided 
that $\zeta \geq \Delta_0$.

As discussed for the two-dimensional case in \cite{tempere}, also in the 
3D case, in the zero-temperature limit, the state of the system can be 
thought of as a superfluid in which the particles with momenta $\left| 
{\bf k} \right| \in \left[ k_-, k_+ \right] $ contribute as normal 
phase particles.

\section{Condensate of Cooper pairs}
\label{sec:uniform}

This section is devoted to study the condensate fraction, that is the ratio 
\beq
\phi \equiv \displaystyle{\frac{N_0}{N}} 
\eeq
where $N_0$ is the condensate number 
of Cooper pairs and $N$ is the total number 
of fermions. $N_0$ is the largest eigenvalue of the two-body density 
matrix \cite{leggett2}, and a finite value of $\phi$ in the thermodynamic 
limit ($N,V\to \infty$ and $n=N/V$ constant) 
implies off-diagonal long-range order 
\cite{penrose}. As a result of spontaneous breaking of $U(1)$ symmetry 
$N_0$ is given by \cite{salasnich1,leggett2}
\beq
N_0 = \sum_{\sigma \sigma'}\int d^3{\bf r} \; d^3{\bf r} \; 
| \langle {\psi}_{\sigma}({\bf r}) {\psi}_{\sigma'}({\bf r}') \rangle |^2 \; ,  
\eeq
and it can be calculated by using the path integral formalism. 
From the Nambu-Gorkov Green function we find 
\beq
N_0 = \frac{1}{\beta^2} \sum_\mathbf{p} \sum_{n} \sum_{m} {G}_{21} 
\left( \mathbf{p}, i \omega_n \right) {G}_{12} 
\left( \mathbf{p}, i \omega_m \right)
\label{eq:n000}
\;,\eeq
where ${G}_{21}$ and ${G}_{12}$ are obtained by inverting Eq. 
(\ref{eq:green1}): 
\beq
{G}_{12} \left( {\bf k}, i \omega_n \right) = -
\frac{\Delta_0}{\left( i \omega_n - \zeta \right)^2 - 
\xi_{\bf k}^2 - \left| \Delta_0 \right|^2 } = {G}_{21} 
\left( {\bf k}, i \omega_n \right)
\;.\eeq

After performing the summation over the Matsubara fermionic frequencies, 
one gets the condensate number (\ref{eq:n000}) as a function of the 
chemical potential $\mu$ (the left formula of Eq. (\ref{hshd})) and of 
the MF order parameter $\Delta_0$  
\beq
N_0 = \sum_{\bf k} \frac{\Delta_0^2}{4 E_{\bf k}^2} \left( \frac{1}{2} 
\tanh \left( \frac{\beta}{2} \left( E_{\bf k} + \zeta \right)\right) + 
\frac{1}{2} \tanh \left( \frac{\beta}{2} \left( E_{\bf k} - \zeta \right)
\right) \right)^2 \; .
\label{eq:mainformula}
\eeq
This formula is the main analytical result of the paper: it gives 
the condensate number of Cooper pairs in term of the energy gap $\Delta_0$, 
average chemical potential $\mu$, imbalance chemical potential $\zeta$, 
and temperature $T$. 
 
We are interested in the zero-temperature analysis of our system. 
In this limit, i.e. $\beta \rightarrow \infty$, the condensate fraction is 
calculated as a momentum space sum. In 
the balanced case the sum runs over all momenta, while for a polarized 
gas the Bogoliubov 
excitations with $\left| {\bf k} \right| \in \left[ k_-, k_+ \right]$ 
do not contribute to the condensate fraction, since the energy of the 
$\downarrow$ fermions $E_{\bf k} - \zeta$ is $\leq 0$:
\beq
\label{eq:n000nT}
N_0=\sum_{\left| {\bf k} \right| \notin \left[ k_-, k_+ \right]} \frac{\Delta_0^2}
{4 E_{\bf k}^2} \; . 
\eeq

\begin{figure}[t]
\begin{center}
{\includegraphics[trim=0 0 0 0,width=12.cm,clip]{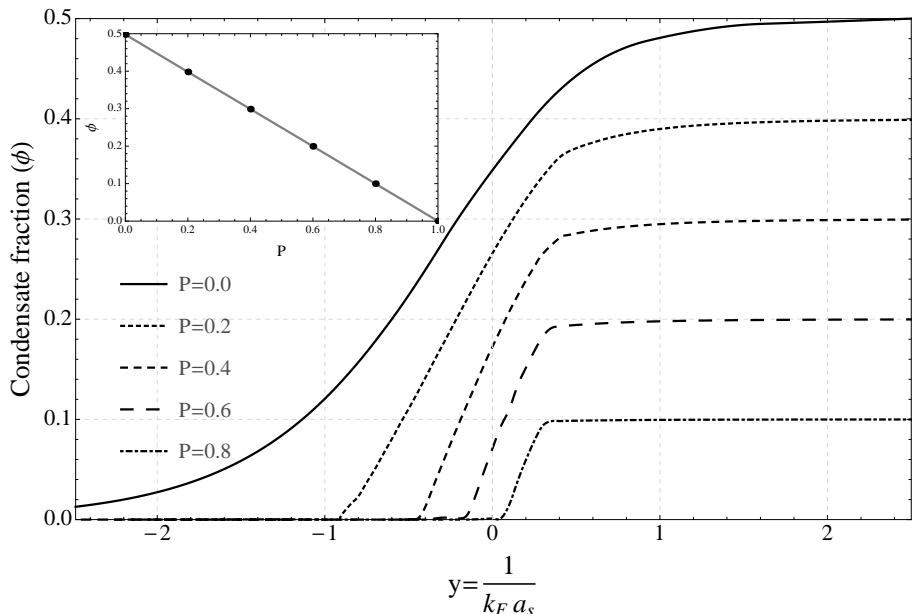}}
\end{center}
\vskip -0.7cm
\caption{The condensate fraction $\phi=N_0/N$ 
as a function of the inverse dimensionless interaction parameter 
$y = 1/(k_Fa_s)$ for different values of the polarization. 
In the inset $\phi$ as a function of the polarization 
$P=(N_{\uparrow}-N_{\downarrow})/(N_{\uparrow}+N_{\downarrow})$ for $y=2$.}
\label{fig:fig1}
\end{figure}

We use this formula for $N_0$ and the zero-temperature 
version of Eqs. (\ref{gapeqT})-(\ref{imbeqT}) to calculate numerically 
the condensate fraction $\phi$ as a function of the dimensionless 
interaction parameter 
\beq
y = {1\over k_F a_s} \; , 
\eeq 
with $k_F=(3\pi n)^{1/3}$ the Fermi wavenumber, and of the polarization 
\beq
P= \frac{N_{\uparrow}-N_{\downarrow}}{N_{\uparrow}+N_{\downarrow}} \; . 
\eeq
The results of these calculations are reported in Fig. \ref{fig:fig1}, 
where we plot the condensate fraction $\phi=N_0/N$ 
as a function of the inverse dimensionless interaction parameter 
$y$ for different values of the polarization $P$. 

In Fig. \ref{fig:fig1} the solid line, corresponding to $P=0$, 
is exactly equal to the solid one reported in Fig. 1 of Ref. \cite{salasnich1}. 
The other curves, corresponding to different finite values of $P$, are instead 
new theoretical results. 
From Fig. \ref{fig:fig1} one observes the expected behavior 
in the deep-BCS regime ($y \ll -1$), 
that is a weak superfluidity which is destroyed as soon as the polarization 
$P$ becomes finite. On the other hand, the results in the 
deep-BEC regime ($y\gg 1$) are easily interpreted considering 
that a number of fermions 
equal to $N_{pairs} = 2\min(N_\uparrow, N_\downarrow)$ 
will produce $N_{pairs}/2$ boson-like bound pairs for sufficiently 
attractive interactions, while the remaining $$N_{normal} = 
N - N_{pairs} = N_\uparrow - N_\downarrow = N P$$ fermions give a 
normal-state Fermi gas. The former kind of particles contribute to the 
condensate fraction, while the latter type does not. Hence, by noting that 
$N_{normal}$ is proportional to the polarization, and by also noting that 
$N_{pairs} = 2 N - 2 N_{normal} $, we expect to observe that in the 
deep BEC regime $N_{pairs} \propto (1-P) \propto \phi$, as verified in 
the inset of Fig. \ref{fig:fig1}. 

\begin{figure*}[t]
\begin{center}
{\includegraphics[trim=0 0 0 0,width=16.cm,clip]{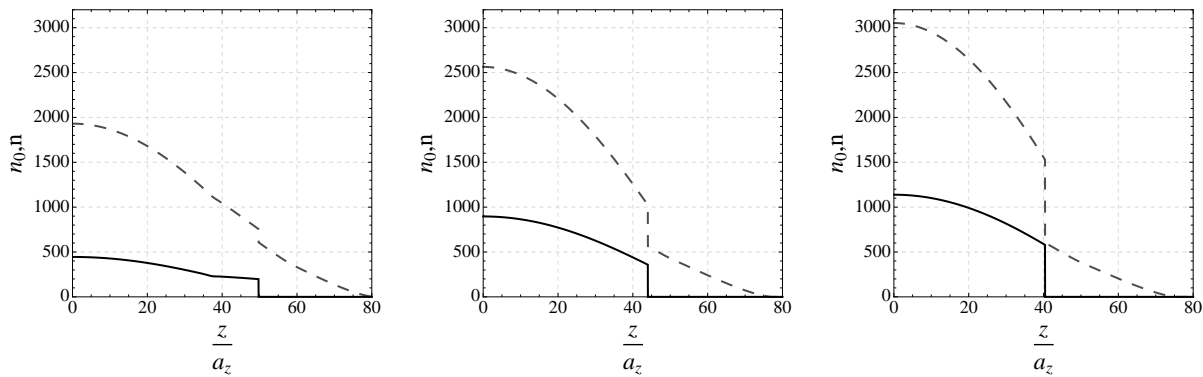}}
\end{center}
\vskip -0.8cm 
\caption{Polarized Fermi gas in the harmonic trap. 
The condensate density profile $n_0(z)$ (solid line)  
and total density profile $n(z)$ (dashed line) in the axial direction $z$ 
for three different scattering lengths. From left 
to right: $y=-0.44$, $y=0.0$, $y=0.11$, where $y=1/(k_Fa_s)$ 
with $k_F=(3\pi^2 n({\bf 0}))^{1/3}$ and $n({\bf 0})$ the total density 
at the center of the trap. Number of atoms $N=2.3 \cdot 10^7$ 
and polarization $P = (N_{\uparrow}-N_{\downarrow})/N= 0.2$. 
Here $a_z =1/{\sqrt{m \omega_z}}$ is the characteristic length 
of the axial harmonic confinement.}
\label{fig:fig2}
\end{figure*}

To conclude this section, we observe that the our 
mean-field approach could be extended by 
determining the magnitude of the contribution from Gaussian fluctuations 
by following \cite{griffin} and thus adding a beyond-mean-field correction 
to the condensate number given by:
\beq
N_0^{(2)} = \frac{1}{\beta} \sum_\mathbf{p, \mathbf{i} \omega_n}
\Tr \left( \tau_3 {G} (\mathbf{p},\mathbf{i} \omega_n) \Sigma 
(\mathbf{p},\mathbf{i} \omega_n) {G}  (\mathbf{p},\mathbf{i} \omega_n) 
\right), 
\eeq 
where $\tau_3$ is the Pauli matrix in the $z$ direction, $\Sigma 
\left( \mathbf{p}, i \omega_n \right)$ is an effective self-energy, 
and the trace is meant to be taken in the Nambu space \cite{griffin}.

As discussed in \cite{griffin}, Gaussian fluctuations  play a key role 
in determining the critical temperature and the finite-temperature properties 
of a Fermi gas. However, in the zero-temperature limit we are investigating 
in the present paper, the effects of Gaussian fluctuations on the condensate 
fraction are not detectable given the experimental sensitivities of the data 
we are comparing our theory with.

\section{External harmonic confinement and local-density approximation}
\label{sec:lda}

Usually a polarized Fermi gas is experimentally investigated 
by confining it in a suitable trapping potential 
$V \left( \mathbf{r} \right)$ given by the axially-symmetric harmonic trap 
\beq
V \left( \mathbf{r} \right) = \frac{m}{2} \left[ \omega_\perp^2 (x^2+y^2) 
+ \omega_z^2 z^2 \right]
\label{trap}
\eeq
where $\omega_\perp$ and $\omega_z$ are the 
transverse and axial trapping frequencies, respectively.

We treat the presence of the external confinement by using the local density 
approximation (LDA). In this approximation, a local chemical potential 
\beq
\mu_{\sigma} \to 
\mu_{\sigma} \left(\mathbf{r} \right) = \mu_{\sigma} - V(\mathbf{r})
\eeq
is introduced and the system is treated as locally 
uniform. As a result, the EBCS gap equation must be solved 
at each point of the 
space for a spatially-dependent gap $\Delta \left( \mathbf{r} \right)$ for 
a given scattering length $a_s$. 

The zero-temperature free energy 
\beq 
F= \Omega_{T=0} 
+\sum_\sigma \mu_\sigma N_\sigma \; , 
\eeq
with $\Omega_{T=0}$ given by Eq. (\ref{omeganoT}),  
exhibits two minima, one in correspondence to a non-zero 
order parameter that we denote by $\Delta_{0} \left( \mathbf{r} 
\right)$ (superfluid phase), and the other corresponding 
to $\Delta_{0}=0$ (normal phase). 
On the interplay between these two minima relies the phase separation in an 
unbalanced trapped Fermi gas. By requiring that the superfluid and the normal 
states have locally the same free energy, one finds a critical value 
$\Delta_{0,c}$ of the energy gap \cite{haque}. 
In the spatial region of the trap where 
$\Delta_0 \left( \mathbf{r} \right) > \Delta_{0,c}$, 
the system is superfluid, while in the spatial region 
where $\Delta_0 \left( \mathbf{r} \right) < \Delta_{0,c}$ the system is normal. 

In Fig. \ref{fig:fig2}, we report the density profile 
$n(z) = n_{\uparrow}(x=0,y=0,z) + n_{\downarrow}(x=0,y=0,z)$ 
in the axial direction $z$, and the condensate density profile,  
$n_0(z) = \int n_0(x=0,y=0,z)$ in the axial direction $z$ 
for three values of the scattering length $a_s$ 
and fixed total number $N=2.3\cdot 10^7$ of atoms and 
polarization $P=0.2$. From the figure one finds that 
the higher-density superfluid regions are located 
at the center of the trap, while the lower-density fully polarized normal 
state are expelled outside (see also \cite{sheehy2}). 
In accordance with other theoretical works \cite{desilva,haque} 
and with experimental data \cite{shin2}, Fig. \ref{fig:fig2} shows 
that the superfluid phase ends abruptly at a critical distance $z_{c}$ 
from the center of the cloud: the condensate fraction jumps 
from a finite value to zero. This effect is a clear manifestation 
of the phase separation. 
Note that Monte Carlo simulations suggest that the LDA 
boundary of the superfluid phase is slightly overestimated \cite{recati,lobo}. 

\section{Interpretation of the results: theory vs. experiment}

\begin{figure}[t]
\begin{center}
{\includegraphics[width=11.cm,clip]{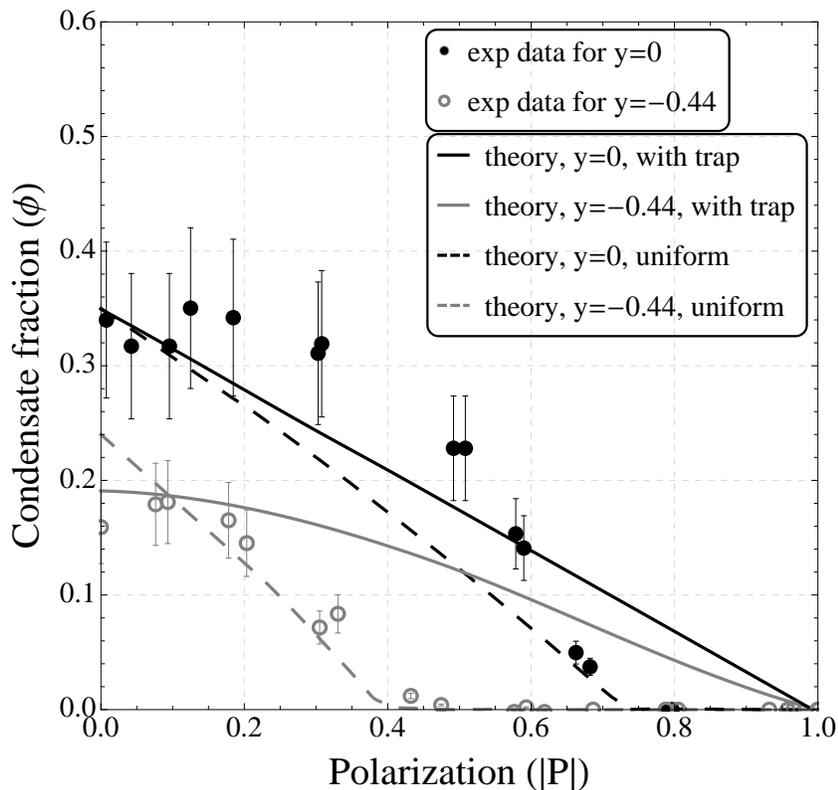}}
\end{center}
\vskip -0.7cm
\caption{Condensate fraction $\phi$ 
as a function of the absolute value of the polarization $|P|$ 
for two values of the dimensionless interaction parameter $y =1/k_F a_s$: 
$y=-0.44$ (open circles) and $y=0.0$ (filled circles). 
Circles with error bars are experimental data 
of $^6$Li atoms taken from Ref. \cite{zwierlein0}. 
Solid lines are our theoretical calculations for the trapped 
system. For completeness, we report also our theoretical 
results of the uniform system (dashed lines).}
\label{fig:fig3}
\end{figure}

The experiment of Ref.~\cite{zwierlein0} with trapped $^6$Li atoms 
addresses the problem of measuring the condensate 
fraction $\phi$ of trapped interacting fermions near the 
unitarity limit as a function of the polarization $P$. 
We attempt a 
comparison of the computed total condensed fraction 
with the experimental data at the lowest
temperature $T= 300$ nKelvin \cite{zwierlein0}. 
In the experiment the scattering length is tuned by means of an external
magnetic tuned across a Feshbach resonance. 
Following Ref.~\cite{bartenstein}, the scattering length $a_s$ as a
function of the magnetic field $B$ near the Feshbach resonance is given by
\beq
a_s = a_b  \left[ 1 + \alpha (B - B_0) \right] 
\left[ 1 + {B_r \over B - B_0 } \right]  \; , 
\eeq
where $B_0 = 83.4149$ mT, $a_b=-1405\cdot 0.53 \cdot 10^{-10}$ m, 
$B_r = 30.0$ mT, 
and $\alpha = 0.0040$ (mT)$^{-1}$. 
The measured condensed fraction at the lowest temperature 
is reported in Fig.~\ref{fig:fig3} as circles and squares 
while our theoretical results are the solid lines. 
For the sake of completeness, we report also the theoretical 
calculations for the uniform system (dashed line)). 

In Fig.~\ref{fig:fig3} it is shown $\phi$ as a function of $P$ 
for two values of $y =1/k_F a_s$:  $y=-0.44$ (open circles) and $y=0.0$ 
(filled circles). The error bars in the experimental data 
for $\phi = \frac{N_0}{N}$ have been calculated 
by using the standard error propagation formula  
considering the $20\%$ error 
on the determination of $N$, as reported in Ref.\cite{zwierlein0}. 
The figure clearly shows that, in the BCS side of the crossover 
up to the unitarity limit ($y \leq 0$), the general trend of the 
experimental data agrees with our EBCS calculation but the agreement 
deteriorates by increasing the polarization $P$. 

It is important to stress that in the BEC regime ($y>0$) the experimental 
data \cite{zwierlein0} are strongly affected by inelastic 
losses and they show a rapid drop of the measured $\phi$ as a function 
of $y$ at fixed polarization $P$. This is clearly incompatible 
(see Fig. \ref{fig:fig1}) with our calculations. For this reason we restrict 
our comparison in Fig. \ref{fig:fig3} to the BCS and unitarity regimes. 

By analyzing our density profiles in Fig. \ref{fig:fig2} for the trapped 
case and the spatial dependence of the the gap $\Delta_0 \left( \mathbf{r} 
\right)$ as defined in section \ref{sec:lda} we can show that the main 
source of disagreement between our theoretical model and the experimental 
data is due to an incorrect modeling of the normal state cloud. 
The boundary of the superfluid region is determined by the condition 
$\Delta_0 \left( \mathbf{r} \right) = \Delta_{0,c}$; as already explained 
in section \ref{sec:lda}, $\Delta_{0,c}$ is determined by comparing the free 
energies of the normal state and of the superfluid state. This quantity 
alone determines the key features of a trapped configuration of a Fermi 
gas and is frequently used in comparing different theories and experimental 
data\cite{zwierlein0}; our model predicts $\Delta_{0,c} \approx 1.4 \zeta$ 
across the whole crossover, while experimental observations and Montecarlo 
simulations\cite{carlson,zwierlein0} at unitarity suggest that the superfluid 
phase breaks down approximately for $\Delta_{0,c} \approx \zeta$.

The origin of this disagreement is better understood by thinking in terms of 
free energy: we model the normal state as a non-interacting Fermi gas; by 
modeling the fermion-fermion interaction in the normal part we would lower the 
energy there, making the normal state energetically favorable in a bigger 
spatial range. By noting that $\Delta_0 \left( \mathbf{r} \right)$ is 
monotonically decreasing as the distance from the center of the trap 
increases, it follows that pushing the phase boundary closer to the trap 
center effectively increases $\Delta_{0,c}$. Therefore we conclude that the 
mean-field approximation of the normal state directly affects the 
polarization values we report. 
On the other hand, the mean-field approximation of the SF core has been 
shown to be much safer: other artifacts could arise as we neglect the 
fluctuation contribution to the condensate fraction in the SF core, but this 
term has been shown to give much smaller contributions to our final 
result\cite{griffin}, well below experimental sensitivities.

At last we take into account the possibility of an 
Fulde-Ferrell-Larkin-Ovchinnikov (FFLO) state., i.e. a grand potential 
minimum competing with the BCS and normal phases, which has been shown to be 
a global minimum for an adequate choice of parameters \cite{fulde,larkin,hu}. 
The existence of such a state in a three dimensional system is still 
subject to intense research: the stability of the FFLO state in a three 
dimensional system has been questioned (see for instance \cite{liao} and 
citations therein), and such a state has not yet been observed in 3D 
systems\cite{devreese}. Eventually, the FFLO state occupies a very 
small region in the parameters space, as reported in \cite{zhang,liao}, and 
would not affect our theoretical estimates significantly larger than the 
other sources of error we analyzed. 
Means of enhancing the FFLO state in 3D which could allow an experimental 
observation have been proposed\cite{devreese,zzheng} on theoretical grounds, 
but up to date they have not been implemented. As the main aim of the present 
paper is to compare our theoretical model with the experimental data in 
\cite{zwierlein}, as the authors in \cite{zwierlein} reported that they 
did not observe a modulation in the condensate density as predicted by the 
theory of the FFLO state, we decided not to include the FFLO state in our 
theoretical description of the unbalanced Fermi gas.

\section{Conclusions}
\label{sec:conclusions}

We have considered a two-component polarized Fermi gas in three dimensions 
across the BCS-BEC crossover. Starting from a path integral formulation, 
we have derived a formula for the condensate fraction at finite temperature. 
In the limit of zero temperature, this system has been analyzed at mean-field 
level (MF), both in the absence and in the presence of an external trapping 
which has been dealt with in the local density approximation (LDA). 
The condensate fraction has been studied as a function of the interatomic 
strength and polarization. We have discussed how an 
enhancement of the population imbalance produces a depletion of the 
condensate fraction which is shown to decrease linearly in the deep BEC 
regime. We have, finally, analyzed the 
harmonically-trapped condensate and the total density 
profiles in correspondence to various 
interaction parameters by commenting that the finite discontinuity between 
the two densities at the phase boundary is a consequence of the phase 
separation. The comparison of our results with the 
experimental data of Zwierlein {\it et al.} \cite{zwierlein0} has shown 
that our predictions are in reasonable agreement with the experimental data 
only in the BCS side of the crossover up to unitarity and for moderate 
polarization values.

\section*{Acknowledgements}

We thank Jacques Tempere, Nicola Manini, and Flavio Toigo 
for fruitful discussions and suggestions. 
GB acknowledges financial support from Italian INFN 
(Istituto Nazionale di Fisica Nucleare). 
LD acknowledges financial support also from MIUR 
(FIRB 2012, Grant No. RBFR12NLNA-002). 
The authors acknowledge for partial support Universit\`a 
di Padova (grant No. CPDA118083), Cariparo Foundation 
(Eccellenza grant 2011/2012), and MIUR (PRIN grant No. 2010LLKJBX).


\section*{References}

\begin{thebibliography}{10}

\bibitem{eagles} D. M. Eagles, Phys. Rev. {\bf 186}, 456, (1969).

\bibitem{leggett} A. J. Leggett, in {\it Modern Trends in the Theory 
of Condensed Matter}, edited by A. Pekalski and J. Przystawa 
(Springer, Berlin, 1980).

\bibitem{nozieres} P. Nozieres and S. Schmitt-Rink, J. Low. Temp. Phys. 
{\bf 59}, 195 (1985).

\bibitem{greiner} M. Greiner, C.A. Regal, and D.S. Jin, Nature (London) 
{\bf 426}, 537 (2003).

\bibitem{regal} C.A. Regal, M. Greiner, and D.S. Jin, Phys. Rev. Lett. 
{\bf 92}, 040403 (2004).

\bibitem{kinast} J. Kinast, S.L. Hemmer, M.E. Gehm, A. Turlapov, 
J.E. Thomas, Phys. Rev. Lett. {\bf 92}, 150402 (2004).

\bibitem{zwierlein} M.W. Zwierlein {\it et al.}, Phys. Rev. Lett. {\bf 92}, 
120403 (2004); M.W. Zwierlein, C.H. Schunck, C.A. Stan, S.M.F. Raupach, 
W. Ketterle,Phys. Rev. Lett. {\bf 94}, 180401 (2005).

\bibitem{chin} C. Chin {\it et al.}, Science {\bf 305}, 1128 (2004); 
M. Bartenstein {\it et al.}, Phys. Rev. Lett. {\bf 92}, 203201 (2004).

\bibitem{ueda} Y. Inada, M. Horikoshi, S. Nakajima, M. Kuwata-Gonokami, 
M. Ueda, and T. Mukaiyama, Phys. Rev. Lett. {\bf 101}, 180406 (2008).

\bibitem{yang} C. N. Yang, Rev. Mod. Phys. {\bf 34}, 694 (1962).

\bibitem{penrose} O. Penrose. Phil. Mag. {\bf 42}, 1373 (1951); 
O. Penrose and L. Onsager, Phys. Rev. {\bf 104}, 576 (1956).

\bibitem{campbell} C. E. Campbell, in Condensed Matter Theories, 
{\bf 12}, 131 (Nova Science, New York, 1997).

\bibitem{salasnich1} L. Salasnich, N. Manini, and A. Parola, 
Phys. Rev. A {\bf 72}, 023621 (2005).

\bibitem{ortiz}  G. Ortiz and J. Dukelsky, Phys. Rev. A {\bf 72}, 
043611 (2005)

\bibitem{monte-carlo} G. E. Astrakharchik, J. Boronat, J. Casulleras, 
and S. Giorgini, Phys. Rev. Lett. {\bf 93}, 200404 (2004).

\bibitem{so1} L. Dell'Anna, G. Mazzarella, and L. Salasnich, 
Phys. Rev. A {\bf 86} 033633 (2011). 

\bibitem{so2} L. Dell'Anna, G. Mazzarella, and L. Salasnich, 
Phys. Rev. A {\bf 87} 053632 (2012).

\bibitem{sala-narrow} L. Salasnich, Phys. Rev. A {\bf 86}, 055602 (2012).  

\bibitem{sala-2d} L. Salasnich, Phys. Rev. A {\bf 76}, 015601 (2007). 

\bibitem{sala-2dlattice} L. Salasnich and F. Toigo, 
Phys. Rev. A {\bf 86}, 023619 (2012).  

\bibitem{sala-beyond} L. Salasnich, P.A. Marchetti, and F. Toigo, 
Phys. Rev. A {\bf 88}, 053612 (2013).  

\bibitem{bedaque} P.-F. Bedaque, H.Caldas,and G.Rupak, 
Phys. Rev. Lett. {\bf 91}, 247002 (2003).

\bibitem{carlson} J. Carlson and S. Reedy, Phys. Rev. Lett. {\bf 95}, 
060401 (2005).

\bibitem{pao0} C.-H. Pao,  S. -T. Wu, and S. -H. Yip, 
Phys. Rev. B. {\bf 73}, 132506 (2005).

\bibitem{son} D.-T. Son and M.-A. Stephanov,Phys. Rev. A {\bf 74}, 
013614 (2005).

\bibitem{miz} T. Mizushima, K. Machida, and M. Ichioka, Phys. Rev. Lett. 
{\bf 94}, 060404 (2005).

\bibitem{sheehy1} D.-E. Sheehy and L. Radzihovsky, Phys. Rev. Lett. 
{\bf 96}, 060401 (2006).

\bibitem{mannarelli} M. Mannarelli, G. Nardulli, and M. Ruggieri, 
Phys. Rev. A {\bf 74}, 033606 (2006).

\bibitem{pieri} P. Pieri and G. C. Strinati, Phys. Rev. Lett. 
{\bf 96}, 150404 (2006).

\bibitem{liu} X.-J. Liu and H. Hu , Europhys. Lett. {\bf 75}, 364 (2006).

\bibitem{hu} H. Hu and X. -J. Liu, Phys. Rev. A {\bf 73}, 051603, (2006).

\bibitem{chien} C. -C. Chien, Q. Chen, Y. He, and K. Levin, 
Phys. Rev. Lett. {\bf 97}, 090402 (2006).

\bibitem{gu} Z.-C. Gu, G. Warner, F. Zhou,  cond-mat/0603091.

\bibitem{martikainen} J. -P. Martikainen,  Phys. Rev. A  {\bf 74}, 
013602 (2006).

\bibitem{iskin2006} M. Iskin and S\'a de Melo, Phys. Rev. {\bf 97}, 100404 
(2006).

\bibitem{desilva} T. N. De Silva and E. J. Mueller, Phys. Rev. A {\bf 73}, 
051602  (2006).

\bibitem{haque} M. Haque and H. T. C. Stoof.  Phys. Rev. A  {\bf 74}, 
011602 (2006).

\bibitem{yi}  W. Yi and L.-M. Duan, Phys. Rev. A {\bf 73}, 031604 (2006).

\bibitem{kinnuen} J. Kinnuen, L. M. Jensen, and P. T\"orm\"a, 
Phys. Rev. Lett. {\bf 96}, 110403 (2006).

\bibitem{parish1} M. M. Parish, F. M. Marchetti, A. Lamacraft, and 
B. D. Simons, Nat. Phys. {\bf 3}, 124 (2007).

\bibitem{sheehy2} D.E. Sheey, L. Radzihovsky, Annals of Physics {\bf 322} 
1790 (2007).

\bibitem{zwierlein0} M. Zwierlein, A. Schirotzek, C.H. Schunck, and 
W. Ketterle, Science {\bf 311}, 492 (2006).

\bibitem{partridge} G. B. Partridge, W. Li, R. Kamar, Y. Liao, and 
R. G. Hulet, Science {\bf 311}, 503 (2006).

\bibitem{zwierlein1} C. H. Schunck, Y. Shin, A. Schirotzek, M. W. Zwierlein, 
and W. Ketterle, Science {\bf 316}, 867 (2007).

\bibitem{fulde} P. Fulde and R. Ferrell, Phys. Rev. {\bf 135}, A550 (1964).

\bibitem{larkin} A. I. Larkin and Yu. N. Ovchinnikov, Sov. Phys. 
JETP {\bf 20}, 762-769 (1955).

\bibitem{tempereft1} J. Tempere, S.N. Klimin, and J.T. Devreese, 
Phys. Rev. A {\bf 78}, 023626 (2008).

\bibitem{tempereft2} S. N. Klimin, J. Tempere, and J.P. A. Devreese, 
J. Low. Temp. Phys. {\bf 165}, 261 (2011).

\bibitem{nambu} Y. Nambu, Phys. Rev. {\bf 117}, 648 (1960); 
L.P. Gorkov, Zh. Eksp. Teor. Fiz. {\bf 36} 1918 (1959). 

\bibitem{tempere} J. Tempere, M. Wouters and J.T. Devreese, 
Phys. Rev. B {\bf 75}, 184526 (2007).

\bibitem{leggett2} A. J. Leggett, {\it Quantum Liquids} 
(Oxford University Press, New York, 2006).

\bibitem{bartenstein}
M.\ Bartenstein {\it et al.}, Phys. Rev. Lett. {\bf 94}, 103201 (2005). 

\bibitem{griffin} N. Fukushima, Y. Ohashi, E. Taylor, A. Griffin, 
Phys. Rev. A {\bf 75}, 033609 (2007).

\bibitem{shin2} Y. Shin, C. H. Schunck, A. Schirotzek, and W. Ketterle, 
Nature {\bf 451}, 689 (2008).

\bibitem{recati} A. Recati, C. Lobo, and S. Stringari, Phys. Rev. A. {\bf 78}, 
023633 (2008).

\bibitem{lobo} C. Lobo, A. Recati, S. Giorgini and S. Stringari, 
Phys. Rev. Lett. {\bf 97}, 200403 (2006).

\bibitem{devreese} J. P. A. Devreese, M. Wouters and J. Tempere, 
Phys. Rev. A {\bf 84}, 043623 (2011)

\bibitem{zhang} W. Zhang and L.M. Duan, Phys. Rev. A {\bf 76}, 042710 (2007)

\bibitem{zzheng} Z. Zheng et al., Phys. Rev. A {\bf 87}, 031602(R) (2013)

\bibitem{liao} Y. Liao et al., Nature {\bf 467}, 567Ð569 (2010)

\end{thebibliography}
\end{document}